\input{aipcheck.tex}

\documentclass[
   a4paper
  ]
  {aipproc}

\layoutstyle{8x11single}
\SetInternalRegister\hbadness{8000}

\newcommand\doingARLO[2][]{%
  \ifx\mmref\undefined #1\else #2\fi
}

\begin{document}

\title
      [MAD-4-MITO]
      {MAD-4-MITO, a Multi Array of Detectors for ground-based mm/submm SZ observations}

\classification{43.35.Ei, 78.60.Mq}
\keywords{Document processing, Class file writing, \LaTeXe{}}

\author{L. Lamagna}{
  address={Experimental Cosmology Group, Dipartimento di Fisica,
  Università di Roma "La Sapienza"\\ P.le A. Moro, 2 - 00185 ROMA (Italy) },
  thanks={}
}

\iftrue
\author{M. De Petris}{
  address={Experimental Cosmology Group, Dipartimento di Fisica,
  Università di Roma "La Sapienza"\\ P.le A. Moro, 2 - 00185 ROMA (Italy) },
  }
\author{F. Melchiorri}{
  address={Experimental Cosmology Group, Dipartimento di Fisica,
  Università di Roma "La Sapienza"\\ P.le A. Moro, 2 - 00185 ROMA (Italy) },
  }
\author{E. Battistelli}{
  address={Experimental Cosmology Group, Dipartimento di Fisica,
  Università di Roma "La Sapienza"\\ P.le A. Moro, 2 - 00185 ROMA (Italy) },
  }
\author{M. De Grazia}{
  address={Experimental Cosmology Group, Dipartimento di Fisica,
  Università di Roma "La Sapienza"\\ P.le A. Moro, 2 - 00185 ROMA (Italy) },
  }
\author{G. Luzzi}{
  address={Experimental Cosmology Group, Dipartimento di Fisica,
  Università di Roma "La Sapienza"\\ P.le A. Moro, 2 - 00185 ROMA (Italy) },
  }
\author{A. Orlando}{
  address={Experimental Cosmology Group, Dipartimento di Fisica,
  Università di Roma "La Sapienza"\\ P.le A. Moro, 2 - 00185 ROMA (Italy) },
  }
\author{G. Savini}{
  address={Experimental Cosmology Group, Dipartimento di Fisica,
  Università di Roma "La Sapienza"\\ P.le A. Moro, 2 - 00185 ROMA (Italy) },
  }
\fi

\copyrightyear  {2001}

\begin{abstract}
The last few years have seen a large development of mm technology
and ultra-sensitive detectors devoted to microwave astronomy and
astrophysics. The possibility to deal with large numbers of these
detectors assembled into multi--pixel imaging systems has greatly
improved the performance of microwave observations, even from
ground--based stations, especially combining the power of
multi--band detectors with their new imaging capabilities.
Hereafter, we will present the development of a multi--pixel
solution devoted to Sunyaev--Zel'dovich observations from
ground--based telescopes, that is going to be operated from the
Millimetre and Infrared Testagrigia Observatory.
\end{abstract}

\maketitle

\section{Introduction}

The Experimental Cosmology Group at the University of Rome "La
Sapienza" has long been active in the development of
instrumentation devoted to microwave astronomy from ground--based
stations and balloon--borne platforms. In particular,
observational campaigns have been performed from the Testa Grigia
Observatory (3500m a.s.l.) searching for Sunyaev Zel'dovich effect
on nearby clusters of galaxies, with special regard for the Coma
cluster (A1656). Thanks to the experience accumulated in dealing
with the site characteristics and the general issues of such
measurements, and due to the importance of the SZ effect as a
cosmological probe and a unique source of information on cluster
physics, the instrument is going to be upgraded from a
single--pixel, 4--band photometer into a multi--band bolometer
array. It is planned to become operational by the end of 2002,
when it will be employed to perform a systematic arcminute--scale
search for the effect over a sample of nearby clusters.
\section{The Sunyaev Zel'dovich effect as a pure estimator on the cosmological distance ladder}
The full power of SZ measurements in providing systematic--free
and redshift independent estimations of cosmological parameters
has been widely explored in recent papers \cite{rephaeli2001}, as
well as demonstrated from a series of experimental results from
interferometric arrays and single dish observations in the radio
region \cite{carlstrom2000}.

SZ effect \cite{suze72} arises from inverse Compton scattering of
CMB photons over a population of hot electrons present in galaxy
clusters atmospheres, and is separated into a {\it thermal}
component, due to the effect of velocity distribution in the hot
gas, and a {\ kinematic} component, arising from the bulk motion
of the hot gas along the line of sight. While the latter has not
yet been object of a systematic search (apart from few
ground--based observational campaigns from the SUZIE group), the
thermal SZ effect has already been observed from a wide variety of
instruments, mainly operating in the radio region with single dish
or interferometric techniques. Combining SZ measurements over a
cluster of galaxies with accurate X--ray surface brightness
information provided from satellite platforms yields an estimate
of the cluster angular diameter distance, so that a sampling of
the redshift--distance Hubble diagram is possible. This can bring
out an estimate of the main cosmological parameters ($H_0$ and
$\Omega_m$) along with information on the total mass of the
cluster. In particular, the possibility to extract the Hubble
constant $H_0$ from a combination of SZ and X--ray data makes this
kind of measurements the natural complement to the mainly
$\Omega$-sensitive CMB power spectrum information.
\subsection{SZ observation program from MITO}
The Millimetre and Infrared Testagrigia Observatory has been
devoted to SZ search for the last 3 years, during which the main
issues of SZ detection (at instrumental as well as at data
reduction level) have been studied and cleared with the aid of a
wide field ($\sim 17'$) single pixel photometer operating in 4
bands, chosen to match the main features of thermal SZ spectrum
with the highest transparency windows of the atmospheric emission
spectrum. The Coma cluster ($A1656$) has been used as a benchmark
for testing instrument and observation strategies (drift--scans
with azimuthal 3--field modulation have been performed recently),
along with software simulation capabilities and data reduction
techniques. The latter, in particular, are based on the
combination of spectral and spatial cross--correlation of sky
signals (which include both CMB data and atmospheric noise) in the
spectral bands \cite{depetris2001,lamagna2001}.


\section{The new MAD-4-MITO}
\begin{figure}
  \centering
  \includegraphics[width=9cm,keepaspectratio]{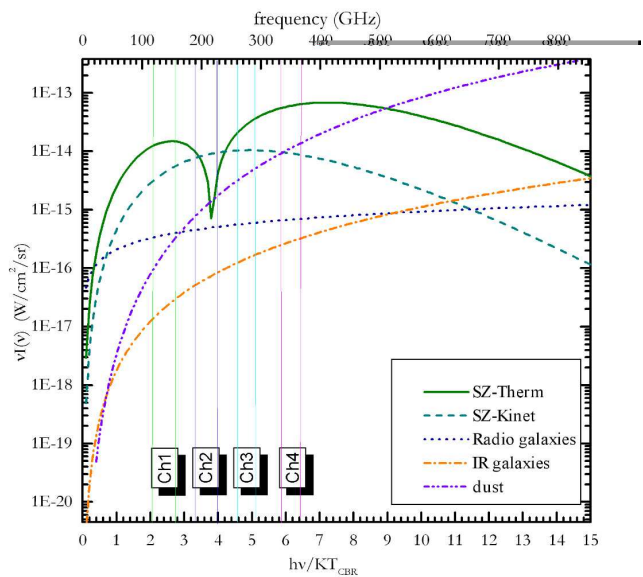}
  \caption{Main astrophysical contributions to signal in the 4 bands selected
  for the MAD experiment.}\label{fig:szcont}
\end{figure}

MAD (Multi Array of Detectors) is an experiment designed to
operate in the 4 MITO bands ($140,220,270$ and $350GHz$). 
These have been chosen to match the best atmospheric transparency
spectral windows and, at the same time, to exploit the full
potential of thermal SZ effect spectral signature (see fig.
\ref{fig:szcont}). The highest frequency channel has been
implemented to monitor the foreground contribution of galactic
dust emission, which becomes more significant with increasing
frequencies and decreasing galactic latitudes. Each channel is
designed to receive radiation from the focal plane after free
propagation in the optics system, and then it pixelizes the focal
plane image into nine regions of $\sim 4$'  dimensions, arranged
in a $9\times 9$ bolometer array (see below). Thus, pixelization
is performed only after splitting the focal plane into 4 identical
spectrally defined beams, instead of employing optically insulated
beam guides directly coupled to a single focal plane image. This
ensures uniform band selection on each pixel, and should provide
good efficiency to the whole system due to the few optical
elements deployed between the telescope focal plane and each
detector array.

\subsubsection{Telescope}
MAD is designed to operate at the $2.6$m telescope of the Testa
Grigia site, located at $3500$m a.s.l. on the Alps, just along the
Italian/Swiss border line. The station, famous for the good
atmospheric conditions and low water vapour content (less than
$1$mm during Winter months, with minimum peaks in
December/January), ensures ideal observing conditions for
microwave experiments, and has been used by the Rome group along
the years as operating base for its main single-pixel SZ-devoted
experiment \cite{depetris1996}, and as a benchmark site for the
Diabolo cryostat \cite{benoit2000}. The telescope has an
aberration-free field of view of $20$', which will be fully
pixelized from the MAD optic system. Recent improvement in the
alignment of the different optical components (especially the
wobbling subreflector) and baffling at the focal plane level,
ensure good control of modulated spurious signals, which may
propagate through the demodulation system as slowly variable
offsets, with time frequencies similar to those of the
astrophysical signal coming from drift--scanned SZ sources, and
thus potentially hard to discriminate on the basis of pure Fourier
analysis.

\subsubsection{Cryostat}
As any other component of the experiment, MAD's cryostat has been
designed to ensure long operation times with minimal servicing,
enabling us to concentrate on observations and treatment of newly
acquired data from calibration and SZ sources while still
performing observations at the telescope.
\begin{figure}
  \centering
  \includegraphics[width=9cm,keepaspectratio]{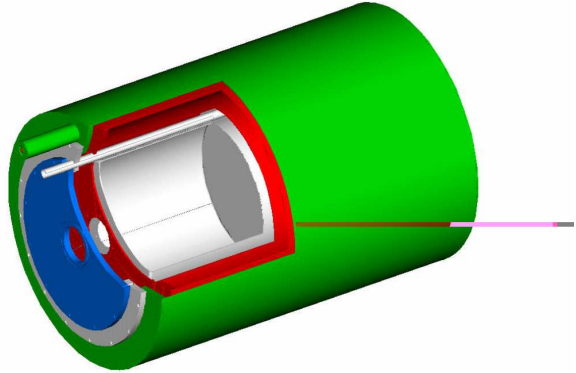}
  \caption{MAD cryostat layout. Short optical paths constrain the detector work area
  in the upper part of the dewar.}\label{fig:cryostat}
\end{figure}

The cryostat (fig. \ref{fig:cryostat}) is designed to have an
upper radiation window, with the work area limited to a few tens
of cm immediately below. This is necessary to assemble the minimum
number of optical elements in front of the detectors and still
have the possibility to operate in 4 different bands: the system
is characterized from extremely small optical paths from the
telescope focal plane to the four detector arrays (fig.
\ref{fig:layout}) and thus needs to be concentrated in the upper
part of the cryostat. This brought to the need of a good
thermalization system, with the helium tank directly below the
optics box, and the full height nitrogen tank providing $77K$
radiation shielding by itself. The goal performance of the
cryostat, whose assembling is still on its way, is to have a fully
thermalized system at $4K$ for $7-8$ days of continuous operation,
without any need for cryogen refilling. With the heat inputs from
radiation window, internal tank supports, and wiring for 40
detectors, it should need about $30l$ for each of the liquid
cryogens to reach this performance. Due to internal layout and
tank shaping, the cryostat still remains very compact, with an
external height of less than $70$cm.
\subsubsection{Two-stage $He^3-He^4$ refrigerator}
Detector cool--down to the operating temperature of $300$mK will
be made possible by means of a two stage $He^3-He^4$ adsorption
refrigerator. Each of the two parts of the system is made of a
self--contained condensation chamber, directly connected to the
cold stage, a heat exchanger, thermally connected with the main
$He^4$ bath, and a cryopumping chamber filled with the proper
amount of active charcoal, which may be thermally controlled using
an externally powered heater and an electro--mechanical heat
switch. The system uses cryopumping on $He^4$ chamber to bring the
cold stage below $He^3$ condensation temperature, then the second
cryopump lowers the liquid $He^3$ (and thus the cold stage)
temperature to the operating level of $\sim 300$mK. The same
refrigerator has been successfully used with the present MITO
cryostat, reaching $290$mK limit temperature and more than $80$
hours of continuous operation with a total heat input of $65\mu W$
on the cold stage \cite{maiani1999}. One of the main advantages of
this design is the total self-consistency of the fridge operation,
since it doesn't need pumping on the main $He^4$ bath with
external devices.
\subsubsection{Detectors and Optics}
MAD's detectors, built at Haller-Beeman, Inc., are standard
composite bolometers made of a sapphire radiation absorber and an
NTD Ge thermistor. They are suspended within a heat trap,
optimized for the 4 operating wavelengths, by means of $30\mu m$
nylon fibers that keep the absorber firm at half-wavelength height
and keep the whole system response fast (expected time constant,
still to measure at this time, is $\tau\simeq 8\div 12 ms$).
Electrical responsivity is about $1.5MV/W$.
\begin{figure}
  \centering
  \includegraphics[width=9cm,keepaspectratio]{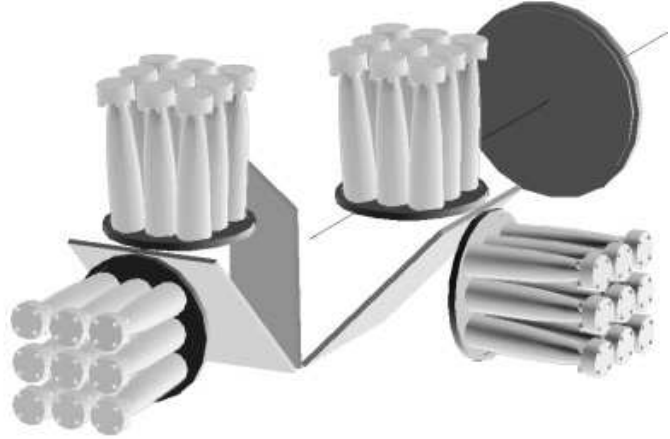}
  \caption{Sketch of detector layout in the MAD optics box. The circular shape indicates
  the position of cryostat radiation window}\label{fig:layout}
\end{figure}

Each of the 4 arrays is arranged in a $3\times 3$ layout (fig.
\ref{fig:layout}), with a tenth bolometer used as a blind monitor
of system noise and channel cross-talk. Since the whole MITO
telescope corrected focal plane area undergoes such pixelization,
single pixel angular resolution is $\sim 4.5'$. The spectral
selection is performed over the unpixelized image, directly below
the focal plane, and the beam is splitted into four images by
means of low pass mesh filters operating at $45°$ incidence.
Single bands are then selected from band-pass filters directly
above the multi--mode Winston concentrators that couple detectors
with the spectrally selected image. We think that this solution
will ensure good optical efficiency to the system, especially if
compared to closed light--pipe solutions, but we are aware of the
problems that can arise from the need for large filters with
uniform performance in order to have pixel--independent spectral
selection. The focal plane layout, provided that the sky image is
kept fixed on each array through the sidereal motion, allows for
continuous detection over the desired direction and the nearest
ones, and, above all, allows for simultaneous observation of the
pixelized sky region in the different bands: this will ensure low
systematics arising from time--delayed signal detection over the
pixels and the different spectral regions.
\subsubsection{Readout electronics and data reduction}
Each detector is DC coupled with a bias circuit and a differential
readout circuit, designed to keep low levels of microphonics and
e.m. spurious signals. Low output impedance is achieved by
bufferizing the two outputs of each detector through a dual JFET
amplifier, set to a common drain ({\it i.e.} unity gain)
configuration. Finally, the signal is amplified from a warm
differential preamplifier board, which sends signals directly to
data acquisition hardware. Proper filtering at lowest frequencies
and anti--aliasing makes it possible to perform data acquisition
at few tens of samples/second with 16 bit resolution. We will
employ a commercial software controlled ADC board, for which fast
monitoring software has been developed. Since the default
observation strategy is based on signal modulation through
wobbling of the telescope subreflector, on-the-fly demodulation
has to be performed in order to extract the information content
from each signal. This feature has been included in the
acquisition software, together with specific self-calibraton
functions, such as offset--monitoring, phase shifting of the
different signals with respect to the modulation reference, and
others still to test on the field. We have also included the
possibility of offline access to modulated data, since we plan to
try many other different observation strategies that would allow
for multi--mode demodulation over the same modulated signal, such
as triangle wave azimuthal scanning or even total power
unmodulated scanning, depending on the final detector performance
and contingent weather conditions that could make it possible to
remove the bulk of atmospheric noise by pixel cross-correlation
alone.


\section{Conclusions}
We have shown the main characteristics of the MAD project, a
multi--pixel $4$--band detector designed for SZ search from
ground--based stations, that will become operational from the
Testa Grigia station within next Winter. The project headlines
have been designed on the basis of the experience accumulated in
the last $3$ years with the single-pixel instrument that is still
operational at the MITO telescope, and has already allowed for SZ
detection over the Coma cluster. With the new instrument, we plan
to perform an extensive observation program to measure the SZ
effect over few tens of clusters and thus map the Hubble diagram
from SZ detection over arcminute scales. This would also bring out
many candidate SZ sources to include in the upcoming OLIMPO
balloon--borne experiment observation program that is undergoing
planning and design in our group.






\begin{theacknowledgments}
This project is being funded from MURST and University of Rome "La
Sapienza". We also wish to thank P. de Bernardis and S. Masi for
continuous support during our work.
\end{theacknowledgments}


\doingARLO[\bibliographystyle{aipproc}]
          {\ifthenelse{\equal{\AIPcitestyleselect}{num}}
             {\bibliographystyle{arlonum}}
             {\bibliographystyle{arlobib}}
          }
\bibliography{2k1bcproc}

\end{document}